# Stimulated Brillouin scattering in dispersed graphene

I. M. Kislyakov,[1,2,*] J.-M. Nunzi,[1,2,3] X. Zhang,[1,2] Y. Xie,[1,2] V. N. Bocharov,[4] and J. Wang[1,2,5,†]

[1] *Laboratory of Micro-Nano Photonic and Optoelectronic Materials and Devices, Shanghai Institute of Optics and Fine Mechanics, Chinese Academy of Sciences, Shanghai 201800, China.*

[2] *Key Laboratory of Materials for High-Power Laser, Shanghai Institute of Optics and Fine Mechanics, Chinese Academy of Sciences, Shanghai 201800, China.*

[3] *Department of Physics, Engineering Physics and Astronomy and Department of Chemistry, Queen's University, Kingston K7L-3N6, Canada.*

[4] *Center for Geo-Environmental Research and Modeling (Geomodel), St. Petersburg State University, St. Petersburg 198504, Russia*

[5] *State Key Laboratory of High Field Laser Physics, Shanghai Institute of Optics and Fine Mechanics, Chinese Academy of Sciences, Shanghai 201800, China*

We address the problem of the interaction of powerful laser radiation with a transparent substance containing a low concentration of strongly absorbing nanoparticles under the condition of Stimulated Brillouin scattering (SBS) whose behavior in such media was unexplored. SBS energies in N-methyl-2-pyrrolidone (NMP) and water are measured at the wavelength of 532 nm. An experimental value $g_B = 18.6 \pm 1.8$ cm GW$^{-1}$ for a previously unknown Brillouin gain factor of NMP is reported. A strong SBS quenching in the liquids by graphene nanoflakes is found. SBS threshold linear dependences on graphene absorption coefficient (concentration) are established and found suitable for the detection of small nanoparticles quantities in water with a minimal detectable concentration of $5 \cdot 10^{-8}$ g cm$^{-3}$. The effect is considered through an electrostriction – thermal expansion antagonism induced by carbon vapor bubbles formation. It is found to be very sensitive to changes in density, refractive index and acoustic absorption coefficient. Using the SBS method, their determination and the bubbling nanosecond time-scaling can be performed. The effect is advantageous for SBS suppression when it is undesirable in laser technologies and optical telecommunication networks.

## I. INTRODUCTION

Two dimensional (2D) nanomaterials (graphene, black phosphorus, transition atoms chalcogenides, etc.) are hottest new trend in science of nanostructures due to their peculiar geometry and electronic spectra determining unordinary properties of electrons and heat transport, as well as broad-ranging possibilities for new nanostructure constructing.[1,2] They also show an impressive gain of nonlinear optical properties like two-photon absorption and absorption saturation[3,4] providing good perspectives of their application in laser techniques, optical informatics and telecommunications.[5]

Ongoing studies of optical phase conjugation in 2D materials are mainly confined to harmonic generation.[6-8] In a recent work[9] the four-wave mixing method was applied to study graphene monolayer nonlinearity in near infrared, and found a large $\chi^{(3)}$ value equal to $2.1 \cdot 10^{-15}$ m$^2$V$^{-2}$ which is *ca.* 7 orders of magnitude greater than in bulk insulators like silica and BK7 glass, 3-5 orders of magnitude greater than in bulk semiconductors like silicon, germanium, chalcogenides of cadmium and zinc, metal oxides, and *ca.* 10 times greater than in thin plasmonic gold films and nanoparticles.[10] Most recently even a larger value of third-order susceptibility: $\chi^{(3)} = 6.3 \cdot 10^{-14}$ m$^2$V$^{-2}$ has been obtained in graphene nanoribbons at frequencies in the middle infrared region close to the transverse plasmon resonance.[11] In the presence of these not yet abundant but impressive advances of phase conjugation in graphene, the effect of 2D materials on stimulated Brillouin scattering (SBS) remains unknown. Meanwhile, this effect is important in laser and fiber telecommunications, and currently attracts theoretical considerations[12-14] concerning bulk and composite semiconductor materials and even practical design work based on them.[15]

These reasons motivate us to experimentally apply the SBS method to 2D nanostructures, and here we report the character of SBS in liquid graphene suspensions. We observed net SBS effect in water and N-methyl-2-pyrrolidone (NMP) and changes in SBS energy caused by the addition of graphene suspensions into the solvents which show a strong quenching of SBS by even vanishingly small concentrations of graphene nanosheets, corresponding to the absorption coefficient by the order of 0.001 cm$^{-1}$ in the case of water. We explain here the effect observed as a result of interference of gratings formed in liquid both by electrostriction and thermal expansion processes. By means of computer simulations of measured concentration dependences of SBS threshold energies we show that the thermal expansion is determined by carbon vapor bubble formation and strongly influences on the refractive index through the density changes and on acoustic wave absorption coefficient through bubble compressibility. This allows to evaluate a scale of effective bubble size in the suspensions under the conditions of powerful laser propagation.

[*] iv.kis@mail.ru
[†] jwang@siom.ac.cn

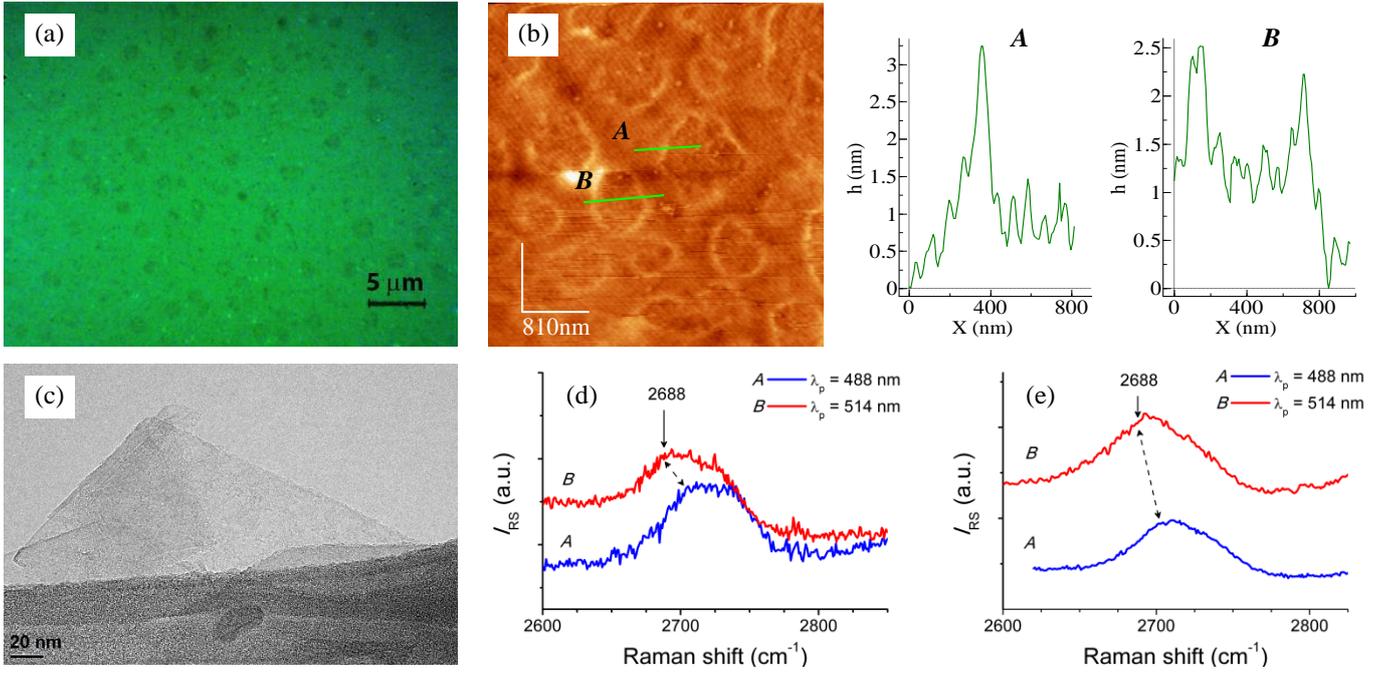

FIG. 1. Optical (a) and AFM (b) images and transverse profiles $A$ and $B$ of Gr$_{water}$. (c) TEM image of Gr$_{NMP}$. Raman G′ band spectra of Gr$_{water}$ (d) and Gr$_{NMP}$ (e): curves $A$ and $B$ correspond to the excitation wavelengths 488 nm and 514 nm correspondingly. More data are presented in Supplemental Material (SM).

The similar effect in suspensions of molybdenum disulfide (MoS$_2$) in isopropanol and of single wall carbon nanotubes (SWCNT) in water has also been observed. In particular, SBS has been detected in the SWCNT suspension at the dilution with absorption coefficient value 0.016 cm$^{-1}$.

## II. MATERIALS AND METHODS

### A. Graphene nanoparticle morphology

Graphene suspensions were prepared from graphite powder by the way like that described earlier.[16] Sodium cholate (NaC) surfactant was used to stabilize the aqueous suspension. Details of suspension preparation and absorption cross-section, $\sigma_e$, measurement are available in Appendix A. We evaluated $\sigma_e = (0.80 \pm 0.16) \cdot 10^4$ cm$^2$g$^{-1}$ of graphene in the aqueous suspension using the density value measured by gravimetric method. Olympus BX53 optical microscope, FEI Tecnai G2 F20 TEM and Bruker Nano Inc. Dimension 3100 AFM were used to visualize graphene. Graphene flakes with transverse sizes from 0.2 to 2 μm were observed in a dried drop of aqueous suspensions (Gr$_{water}$) on a glass plate both in the optical [Fig. 1(a)] and in the atomic force [Fig. 1(b)] microscopes. We could not obtain the same pictures of graphene from NMP (Gr$_{NMP}$), where separate flakes probably strongly aggregate during drying. In the case of aqueous suspension, it turned out apparently possible because the surfactant prevents graphene aggregation by forming a sort of coat [high spikes in Fig. 1(b) at the flake edges]. The surfactant can also improve the visibility of graphene sheets, and this can be the alternative reason why we can see the flakes dried from water and do not see them from NMP. However, relying on the transmission electron microscopy images obtained [see Fig. 1(c) and SM], we believe that graphene flakes in NMP are of similar sizes.

The flakes thickness measured by AFM [see Fig. 1(b) and SM] varies from 0.3 nm to 1.5 nm approximately, evidencing the few layers morphology with middle thickness around 1 nm corresponding to 3-layer graphene whose thickness is estimated as a sum of two 0.34 nm interlayer distances and two 0.17 nm van der Waals radii. The suspensions were further studied by Raman spectra analysis. The Raman studies were performed at the Research park of St. Petersburg State University, center for Geo-Environmental Research and Modeling "Geomodel", using a Horiba Jobin-Yvon LabRAM HR800 spectrometer equipped with a microscope Olympus BX-41. Spectral resolution provided was better than 3 cm$^{-1}$. Raman spectra were excited by an Ar$^+$ laser at 514 nm and 488 nm wavelengths with maximal power of 5 mW at the sample. A double Raman resonance G′ mode band of graphene dried from the aqueous suspension [see Fig. 1(d)] shows the spectral position and shape (a complex asymmetric band with maximum at the Raman shift of *ca.* 2688 cm$^{-1}$, and spectral width of *ca.* 60 cm$^{-1}$) which also correspond to the 3-layer graphene.[17] Graphene dried from NMP suspension [Fig. 1(e)] shows an intermediate shape of its G′ mode band between 2-layer and 3-layer graphene band shapes. Based on these data we suppose the preponderance of 3-layer graphene flakes in our suspensions.

### B. Nonlinear optical setup

SBS observations were performed at room temperature using the second harmonic ($\lambda$ = 532 nm) of 4 ns pulsed Continuum Minilite II Nd:YAG laser operating at 1Hz pulse repetition frequency.

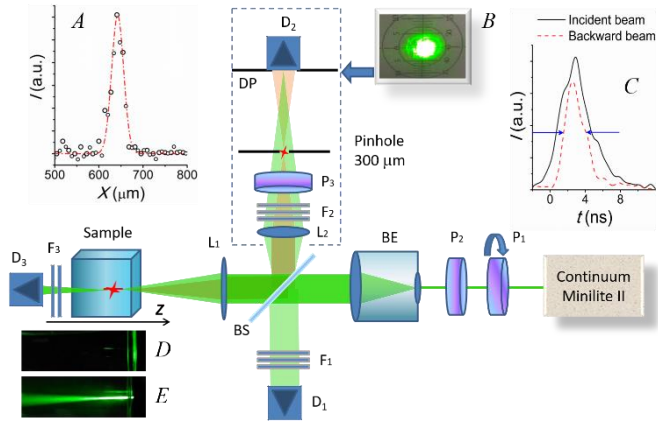

FIG. 2. Schematic of the SBS set up: Continuum Minilite II – laser source; $P_1$, $P_2$, $P_3$ – polarizers; BE – beam expander; BS – beam splitter; $L_1$, $L_2$ – lenses; $F_1$, $F_2$, $F_3$ – neutral filters sets; $D_1$, $D_2$, $D_3$ – detectors; S – sample cuvette; PH – pinhole; DP – diaphragm; Sc – screen; dashed rectangle shows a light protection mask. Insets: $A$ – X-axe beam intensity distribution, measured at $L_2$ focal position (solid) and its Gaussian fit (dash); $B$ – SBS image on the screen from pure NMP; $C$ – time profiles of incident beam (solid) and backward scattered (dash); $D$ – image of the interaction area from the cuvette with pure solvent; $E$ – image of the interaction area from the cuvette with a pristine graphene suspension (no SBS observed).

The setup is schematically shown in Fig. 2. Cross polarizers $P_1$ and $P_2$ were used to change the laser pulse energy. A beam expander telescope (BE) was used to decrease the beam divergence angle in 5 times from its initial level of $\theta_0 = 1.8$ mrad (measured by a ruler at the 4 m base). The beam was focused in a 3 cm path quartz cuvette by 150 mm biconvex lens $L_1$. The beam spatial distribution after the lens being close to Gaussian in the focal point (see the inset $A$ in Fig. 2), is influenced by diffraction at other $Z$ positions. Z-scan study of the beam spot radius ($W$) performed in the air and approximations of its results by the function of the form $W(Z) = W_0(1+(Z/Z_R)^2)^{1/2}$ gave the waist radius $W_0 = 41 \pm 7$ μm and the Rayleigh length $Z_R = 1.0 \pm 0.2$ mm.

The backward scattered beam was reflected by the beam splitter BS and observed on the screen put after 300 μm pinhole PH. Inset $B$ in Fig. 2 shows a visual picture of the SBS effect which has specific threshold behavior. The backward beam polarization followed the polarization of the incident radiation. Time profiles measurements performed with Thorlabs DET 025AL/M 2 GHz silicon detector manifest a pulse-time reduction from $\tau_p = 3.7 \pm 0.3$ ns (FWHM) in the incident beam to $\tau_p = 2.6 \pm 0.2$ ns in the backward beam (inset $C$ in Fig. 2) that gives evidence of its nonlinear origin.

Energies of incident, back-scattered and transmitted beams were measured by silicon detectors Thorlabs PDA100A-EC calibrated both by Coherent J50MB-YAG-1561 and by Thorlabs ES111C pyroelectric heads to take into account possible systematic error related to the energy meters. The difference in the calibrations was within 5%. The maximal SBS beam energy was found to be *ca.* 0.9 mJ in organic liquids and *ca.* 0.05 mJ in water at the ultimate achievable input energy 4.5 mJ. No SBS was detected in pristine (not diluted) suspensions of graphene nanosheets in water and NMP. Then, graphene suspensions were added into pure solvents by small portions, evaluating its linear absorption coefficient $\alpha_e$ (exponent index in Lambert-Beer law: $\alpha_e = \alpha_{10} \cdot \ln(10)$, where $\alpha_{10}$ is instrumentally obtained absorption coefficient) by the dissolution ratio.

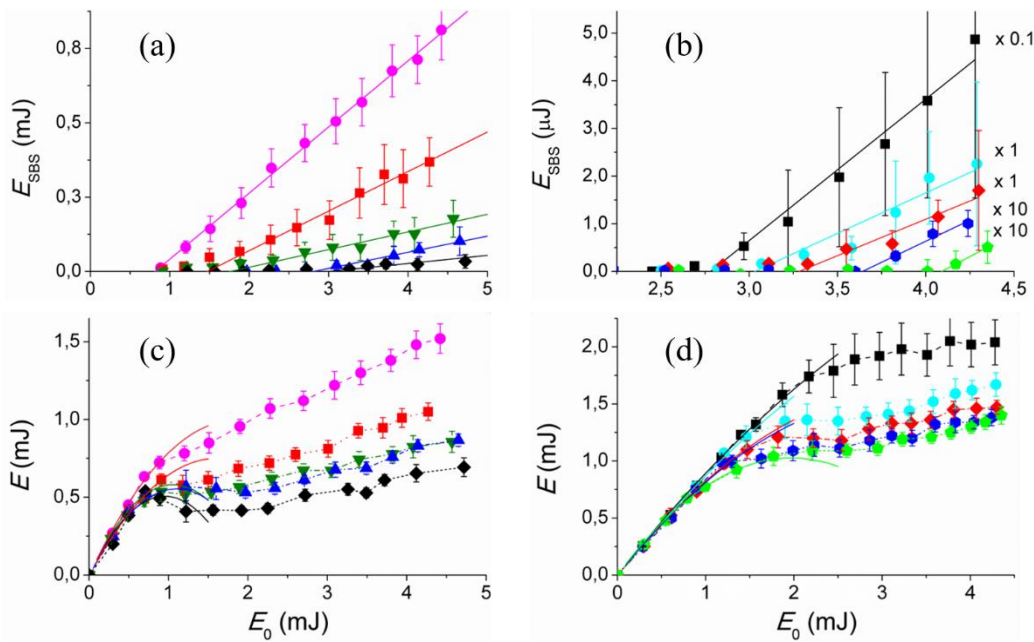

FIG. 3 Back scattered (a), (b) and transmitted (c), (d) beam energies from/through NMP-based (a), (c) and water-based (b), (d) suspensions against incident beam energy; Absorption coefficients $\alpha_e$: (a) and (c) – ●, pure NMP; ■, 0.013 cm$^{-1}$; ▼, 0.023 cm$^{-1}$; ▲, 0.032 cm$^{-1}$; ◆, 0.048 cm$^{-1}$; (b) and (d) – ■, pure water; ○ (0.5 wt.% NaC solution); ◆, 0.001 cm$^{-1}$; ●, 0.002 cm$^{-1}$; ●, 0.004 cm$^{-1}$.

## III. RESULTS OF NONLINEAR OPTICAL MEASUREMENTS

### A. Stimulated Brillouin scattering

SBS beam energy $E_{SBS}$ was measured vs. pump laser energy $E_0$ simultaneously with energy of the beam transmitted through the sample, $E$, versus $E_0$ (optical limiting curves) for different linear absorption coefficients. Most of the dependences obtained are shown in Fig. 3. SBS threshold energies ($E_{thr}$) can be seen on these dependences and numerically found as X-intercepts of linear fittings as shown in Figs. 3(a) and 3(b). They determine the SBS gain coefficient $G$ according the condition:[18]

$$G = \frac{g_B E_{thr} L}{\pi W_0^2 \tau_p} = 25, \quad (1)$$

which corresponds to the regime where spontaneous scattering is amplified to SBS. Here $g_B$ is the Brillouin gain factor, $L$ – the interaction length which in our experiment can be considered as $2Z_R n$, $n$ – refractive index.

Small additions of graphene suspensions in pure solvents remarkably decrease the SBS beam energy and increase its threshold. This is quite opposite of the SBS threshold reduction[14] calculated for a graphene-clad tapered silica fiber where graphene modulates the hypersound wave propagating through the fiber due to an appropriate ratio of photo-acoustic and electro-optic parameters of the core and the clad (especially, acoustic velocity, Brillouin frequency and refractive index). In our case, we see a manifestation of another effect, taking place on the background of acousto-optical modulation of the bulk matter by dispersed graphene nanoparticles and apparently originated in light absorption by them.

The obtained concentration dependence of the threshold energy revealed rather a linear character. Figures 4(a) and 4(b) show the dependences of $E_{thr}$ on $\alpha_e$, which look quite linear both in NMP and in water. Experimental errors in Fig. 4 are within the point's size except the last point where error is large because of very small SBS energy values. The SBS threshold in water was shifted up even with the NaC addition, however, our measurements performed with higher concentrations of NaC (from 1 to 5 wt.%) did not reveal further threshold increasing. Since the NaC concentration used in our suspension is in proximity of critical micellar concentration:[19] 0.68 wt.%, we consider the micellation as the reason of this SBS threshold change through an additional extinction both of optical and acoustic waves. It was extracted from the experimental points [see hollow squares in Fig. 4(b)] and only the effect connected with graphene was further considered.

We see from the Fig. 4 that SBS threshold is very strongly affected by graphene quantities which can hardly even be detected spectrometrically ($\alpha_{10} < 0.01$), and which can be considered as impurities. Thus, the SBS threshold measurement can be considered as a method for such impurity detection in pure liquids. The minimal detectable graphene concentration, $\rho_{Gr\,min}$, can be evaluated through the uncertainty of SBS energy at lower concentrations, $\Delta E_{thr}$, which is biggest in case of water: $\Delta E_{thr} = 0.11$ mJ. The minimal detectable graphene absorption is $\alpha_{e\,min} = \Delta E_{thr}/\text{Slope} = 4\cdot10^{-4}$ cm$^{-1}$, and $\rho_{Gr\,min} = \alpha_{e\,min}/\sigma_e = 5\cdot10^{-8}$ g cm$^{-3}$. The value for NMP can be even smaller because $\alpha_{e\,min}$ consists only $2\cdot10^{-5}$ cm$^{-1}$, however, since the absorption cross-section for the NMP-based suspension is not determined we cannot quote it here.

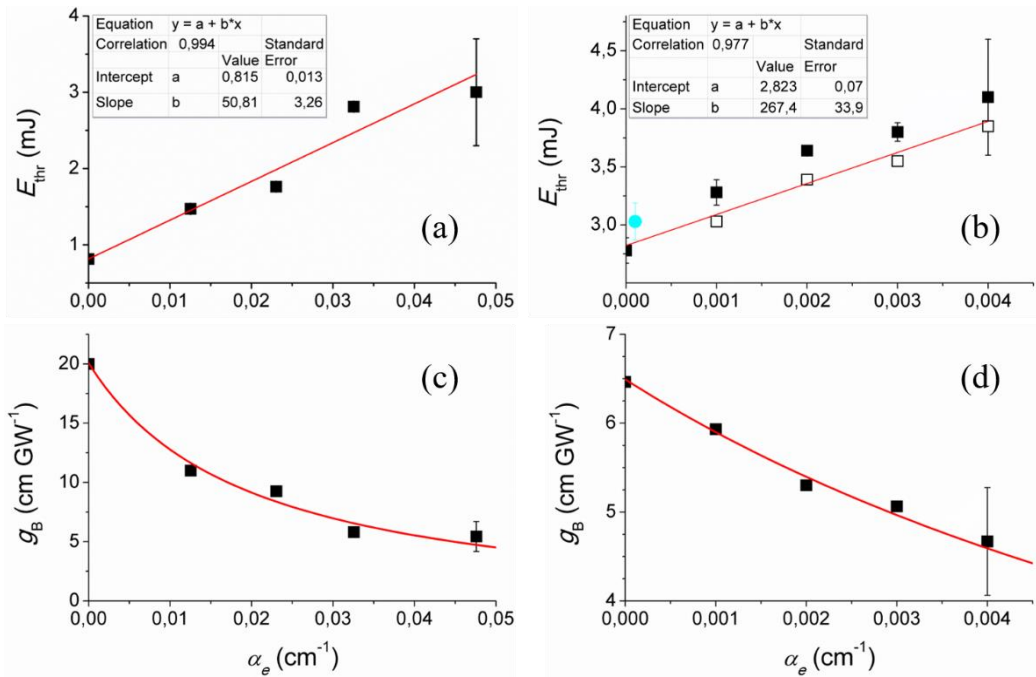

FIG. 4. Dependences of SBS threshold energies (a), (b) and SBS gain factors (c), (d) on linear absorption of graphene suspensions in NMP (a), (c) and in water (b), (d); filled points: experimental results; empty points in (b): correction for the shift due to NaC (blue circle); lines in (a) and (b): linear fitting; curves in (c) and (d): simulations with simultaneous variations of $\rho$ and $\Gamma_B$.

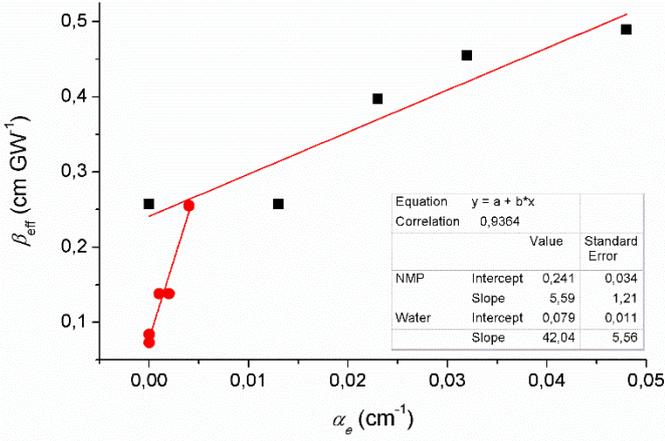

FIG. 5. Dependence of effective NLA coefficients of suspensions on linear absorption of graphene.

The Brillouin gain factor value obtained from the intercept of $E_{thr}(\alpha_e)$ fitting line by means of Eq. (1), $g_B(0) = 6.4 \pm 0.1$ cm·GW$^{-1}$ for water, is in accordance with a theoretical one 6.4 cm·GW$^{-1}$, and however by 35% greater than an experimental one 4.8 cm·GW$^{-1}$.[20] Although we could not find any published value of $g_B$ for NMP, the plausible result for water convinces us in correctness of our value $g_B(0) = 18.6 \pm 1.8$ cm·GW$^{-1}$. It should be noted that the uncertainties indicated for $g_B(0)$ are obtained from the spread of $E_{thr}$ experimental points and do not include systematic errors. An analysis of the latter shows that the error of beam waist $\Delta W_0$ dominates among them in Eq. (1) and consists 34%.

Since the $E_{thr}(\alpha_e)$ dependence is linear, the $g_B(\alpha_e)$ one evaluated using Eq. (1) shows hyperbolic character [points in Figs. 4(c) and 4(d)].

### B. Optical limiting

Figures 3(c) and 3(d) show the optical limiting curves manifesting moderate nonlinear behavior at the energies below SBS thresholds (< 1 mJ for NMP and < 2 mJ for water) which has been considered as effective two-photon absorption (TPA) and fitted by parabolic functions shown in the figures as solid curves:

$$E(E_0) = T_{lin}\left(E_0 - \frac{\beta_{eff} L}{\pi W_0^2 \tau_p} E_0^2\right), \quad (2)$$

where $T_{lin}$ – linear transmittance, $\beta_{eff}$ – TPA coefficient, or effective nonlinear absorption (NLA) coefficient in general case. Results of this approximation are given in the Table I.

We plotted the obtained coefficients against the linear absorption ones and fit by straight lines as shown in Fig. 5. The line slope determines the nonlinear-to-linear absorption cross-sections ratio:

$$\frac{\beta_{eff}}{\alpha_e} = \frac{\sigma_{eff}}{\sigma_e}. \quad (3)$$

TABLE I. Effective NLA coefficients of liquid systems with different content of graphene obtained from the fitting of experimental points by Eq. (2)

| NMP | | Water | |
|---|---|---|---|
| $\alpha_e$, cm$^{-1}$ | $\beta_{eff}$, cm GW$^{-1}$ | $\alpha_e$, cm$^{-1}$ | $\beta_{eff}$, cm GW$^{-1}$ |
| 0 (pure NMP) | 0.257 | 0 (pure water) | 0.073 |
| 0.013 | 0.257 | 0 (NaC sol.) | 0.084 |
| 0.023 | 0.397 | 0.001 | 0.138 |
| 0.032 | 0.455 | 0.002 | 0.138 |
| 0.048 | 0.489 | 0.004 | 0.255 |

The NLA cross-section for the suspension in NMP is found to be $\sigma_{eff} = (5.59 \pm 1.21)$ cm$^2$ GW$^{-1} \cdot \sigma_e$, and for that in water: $\sigma_{eff} = (42.0 \pm 5.6)$ cm$^2$ GW$^{-1} \cdot \sigma_e$. The data on TPA coefficients of pure solvents in visible range are scarce. For water at $\lambda = 532$ nm we have just an upper bound of it: $\beta_{eff} < 0.004$ cm GW$^{-1}$, found in picosecond time regime.[21] Our value is 20 times greater. Hence, our value comprises other processes, enhancing from picosecond to nanosecond time scale. In view of absence of any linear transmission in the visible range, we can exclude reverse saturable absorption effects related to long-live states and consider the light scattering on augmenting density fluctuations as the most realistic reason. These fluctuations follow the intensity fluctuations (speckles) due to the electrostriction forces even prior to the visible appearance of SBS.[22] At higher energies, around SBS thresholds the limiting curves in Figs. 3(c) and 3(d) show pits and subsequent linear behavior signaling additional reduction of the transmitted energy by SBS whose energy linearly depends on $E_0$ within our experiment accuracy.

On the other hand, for a graphene suspension in NMP with $\alpha_e = 0.223$ cm$^{-1}$ we know a value of $\beta_{eff} = 0.29$ cm·GW$^{-1}$ obtained from Z-scan measurements at $\lambda = 532$ nm, $\tau_p = 4$ ns.[23] That should give us $\beta_{eff} \sim 0.03$ cm·GW$^{-1}$ for our suspensions with ten times smaller absorption. Our $\beta_{eff}$ value for the graphene suspension in NMP: $\beta_{eff}(\alpha_e = 0.023) - \beta_{eff}(\alpha_e = 0) = 0.14$ cm·GW$^{-1}$ is about 5 times greater. Therefore, we think that the linear concentration dependence in Eq. (3) may be saturated at higher concentration of graphene nanoparticles due to aggregation effects (flakes size redistribution).

## IV. THEORY OF THE EFFECT AND SIMULATIONS

### A. Conceptual issues

Propagation of the SBS beam intensity $I_S$ in the case of a weakly absorbing substance is described by the equation:[24]

$$\left(\frac{d}{dz} - \alpha_e\right)I_S = -g_B I_S I_0, \quad (4)$$

where $I_0$ is the incident beam intensity. A graphene addition can change the absorption coefficient of the substance in the left side of the equation. It decreases the Brillouin gain in Eq. (1) by the term $-\alpha_e L$, whose greatest value in our case is achieved in the NMP-based suspension: $\alpha_e L = 0.048$ cm$^{-1} \cdot 0.294$ cm $= 0.014$ and consists only 0.056% of the threshold gain $G = 25$ which has no influence on the effect observed. The effective NLA coefficient decreases the Brillouin gain factor explicitly: $\Delta g_B = -\beta_{eff}$. This quantity is proportional to $\alpha_e$ as we see from Eq. (3), but the ratio $\beta_{eff}/\alpha_e$ values we obtained is much less than the derivatives of the curves in Figs. 4(c) and 4(d): $dg_B(0)/d\alpha_e = -g_B(0)\cdot\text{Slope}/E_{thr}(0) \approx -670$ cm$^2$GW$^{-1}$ (NMP) and $dg_B(0)/d\alpha_e \approx -1500$ cm$^2$GW$^{-1}$ (water). Therefore, no linear losses in Eq. (4) can explain the effect of SBS quenching by graphene.

Somehow analogous effect was observed and described by McEwan and Madden[25] in experiments on degenerated four-wave mixing in carbon black suspensions. They have shown that bubbles induced by laser light whose energy is absorbed by carbon nanoparticles and released as a local temperature rising, form in the substance a transient grating correlated to the interference maxima of two incident laser beams. The fourth beam appears as a result of diffraction of the third beam on the grating. Graphene nanoparticles being good absorbers should form a similar grating from their thermally expanded microenvironment. The difference in our case is that this grating corresponds to the maxima of the incident beam intensity.

The sketch in Fig. 6 shows the processes taking place in the graphene suspensions under the conditions of SBS generation. First, electrostriction forces (red arrows) compress the matter and form the density peak (hence, the refractive index peak) grating which is responsible for the backward scattered wave amplification (the conventional SBS generation mechanism). At the same time, graphene nanoparticles form a "negative" grating of refractive index dips due to thermal expansion which is coherent to the first grating and destroys it.

This background requires a consideration of the thermal expansion process that leads to an addition appearance to the Brillouin gain factor in the form like for thermal-induced scattering but opposite in sign:[24]

$$g_B(\alpha_e) = g_B^e - g_B^T(\alpha_e), \quad (5)$$

where (with electric constant $\varepsilon_0$ for S.I. units):

$$g_B^e = \frac{(\gamma_e \omega_0)^2}{\varepsilon_0^2 c^3 n \rho v_{ac} \Gamma_B} \quad (6)$$

is the electrostrictive part of the gain factor, and

$$g_B^T(\alpha_e) = \frac{\gamma_e \gamma_T(\alpha_e) \omega_0^2}{\varepsilon_0^2 c^3 n \rho v_{ac} \Gamma_B} \quad (7)$$

is the thermo-expansive part.

Notations are: $\gamma_e = \varepsilon_0 \rho (\partial \varepsilon / \partial \rho)_S$ – electrostrictive coefficient; $\gamma_T(\alpha_e) = \varepsilon_0 c^2 \beta_T v_{ac} \alpha_e / (C_p \omega_0)$ – thermo-optic coupling coefficient; $\beta_T = (\partial V/\partial T)_\rho / V$ – thermal expansion coefficient; $C_p$ – specific heat capacity at constant pressure; $\rho$ – solvent density; $S$ – entropy; $n$ – refractive index.

$$\Gamma_B = \frac{(\omega_0 - \omega_B)^2 \eta}{v_{ac}^2 \rho} \quad (8)$$

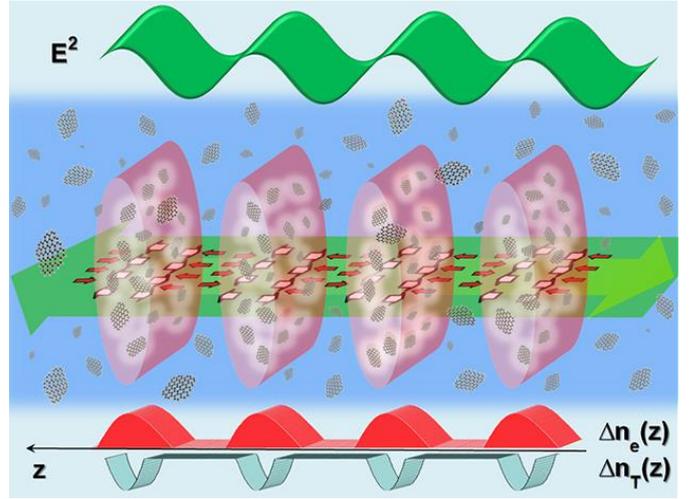

FIG. 6. Sketch of the electrostriction – thermal expansion antagonism occurring in a graphene suspension upon its irradiation and responsible for the SBS quenching.

is the Brillouin line width (acoustic wave dumping coefficient); $\omega_0$, $\omega_B$, – incident laser and Brillouin waves circular frequencies correspondingly; $v_{ac}$ – acoustic wave velocity; $\eta = \eta_s(4/3 + \eta_b/\eta_s)$ – total viscosity of the solvent: $\eta_s$ – shear viscosity, $\eta_b$ – bulk viscosity.

### B. Local heating of graphene flakes

A local temperature rise $\Delta T$ of an absorbing graphene flake drives the process, and the understanding of its scale is important. It can be upper estimated using the simple heat capacity definition relation: $\Delta T = \Delta Q_{abs}/(m_{Gr} \cdot C_p)$, where $\Delta Q_{abs}$ is the heat amount obtained by the graphene flake from the radiation absorbed, $m_{Gr}$ is graphene flake mass, $C_p = 0.7$ J·K$^{-1}$g$^{-1}$ for graphene.[26]

Following the Lambert-Beer law at low absorption and neglecting the scattering: $\Delta Q_{abs} = E_0 F(t) \sigma_e m_{Gr} L/V$, where $V$ is irradiated volume of suspension, $F(t) = \int_0^t I_0(\tau) d\tau / E_0$ – time dependent radiation dose, which can be easily calculated from the laser beam time profile (see Fig. 2, inset C). Considering the irradiated volume is that inside the caustic surface: $V = 8\pi W_0^2 Z_R/3$ for the Gaussian beam, we finally obtain that the temperature rise does not determined by the flake mass or size, but only by the absorption cross-section:

$$\Delta T(t) = \frac{3 E_0 F(t) \sigma_e}{4\pi W_0^2 C_p}. \quad (9)$$

The calculated temporal dependence for the case of input laser pulse energy 1 mJ which approximately corresponds to the minimal SBS threshold is shown in Fig. 7 by solid line.

For the first $\tau_1 = 1.34$ ns a temperature rise of 4830 K is achieved providing the temperature value which approximately corresponds to the graphite sublimation temperature reported by McEwan and Madden.[25] Subsequent sublimation can be rated by:

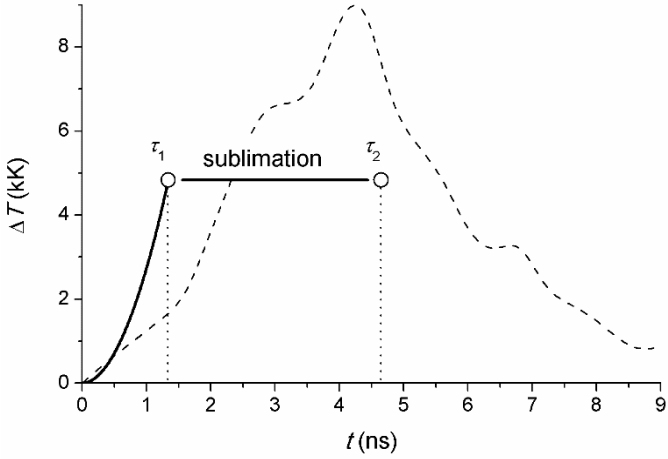

FIG. 7. Temperature rise of a graphene flake upon 1 mJ laser pulse absorption (solid curve); laser beam temporal profile (dashed curve) in arbitrary units.

$$F(\tau_2) = F(\tau_1) + \frac{4\pi W_0^2 \Lambda}{3E_0 \sigma_e},$$

where $\Lambda$ – heat of vaporization, from where the sublimation ending time $\tau_2 = 4.65$ ns has been determined. It should be noted that the presented values are upper estimations. We used parameters known for graphite including the heat of vaporization value $\Lambda = 169.7$ kcal/mole from JANAF thermochemical tables.[27] However, it is arguable that thin graphene flakes can be evaporated more easily. For example, a study of graphite filaments reported in Ref. 27 reveals their effective evaporation at remarkably lower temperatures (3200-3500 K) and with a lower heat of evaporation value: 102.7 kcal/mole. This trend would decrease the evaporation time in our case which would lead to an even stronger affection on the laser pulse propagation.

Carbon atom and its ions have no remarkable absorption lines at $\lambda = 532$ nm and 266 nm, therefore further growth of the temperature after the sublimation can be determined either by carbyne forms in the carbon vapor or by plasma electrons. The plasma is apparently getting generated in the bubbles due to a partial oxidation of the carbon atoms. In both cases, the further temperature rise is hardly evaluable in the framework of the present study.

Therefore, at laser pulse energy around the SBS threshold in NMP, all dispersed graphite nanoparticles become evaporated after 4.65 ns. The characteristic time of SBS is the inverse acoustic wave dumping coefficient ($2\pi/\Gamma_B$), and for most of the studied liquids it is equal to 3-4 ns which means that the graphite evaporation at the near-threshold irradiation is close to interfere with the SBS process. For greater pulse energies this time scale is shorter. Hence, in the presence of graphene flakes, SBS is developing in a medium which represents an emulsion of carbon vapor bubbles and has non-steady temperature fields. Thermodynamic characteristics of such medium can significantly differ from those of pure liquid. The increasing of graphene concentration will also increase the bubble concentration and the material characteristics will change drastically.

### C. Simulation of $g_B(\alpha_e)$ dependence

Only thermo-expansive part of the Brillouin gain factor [see Eq. (7)] shows an explicit dependence on $\alpha_e$, which, however, does describe neither the derivative $dg_B(\alpha_e)/d\alpha_e$ observed, nor the hyperbolic form of the $g_B(\alpha_e)$ dependence. Thus, we must assume an influence of graphene flakes on SBS parameters in Eq. (6) and Eq. (7).

We performed simulations of $g_B(\alpha_e)$ by small (less than 20%) variations of each parameter $\delta P$ at fixed values of all others, referring it to the observed scale of the absorption coefficient change $\Delta \alpha_e$:

$$\delta P = P(1 \pm \Delta_\alpha P \cdot \Delta \alpha_e),$$
$$\Delta_\alpha P = \frac{1}{P}\frac{dP}{d\alpha_e}. \quad (10)$$

Here the value $\Delta_\alpha P$ is an absorptive increment (or decrement) of the parameter. Having varied refractive index, density, acoustic velocity, Brillouin line width and thermo-optic coupling coefficient, a comparative examination of the efficiency of each parameter on the Brillouin gain factor reduction has been made. This approach implies the priority of small variations of the substance characteristics in the observed changes of the net effect, and its reliability is based on numerous observations of the relatively small changes in refractive index,[4,28] acoustic velocity,[29,30] thermal expansion coefficient[31] and frequency shifts, $\omega_0 - \omega_B$,[29] occurring at high power laser action on nano-carbon species in suspensions, as well as on the moderate character of known temperature and pressure dependences of density, speed of sound, viscosity, refractive index, and thermal expansion coefficient of pure NMP[32-34] and water[32,35].

Some of the parameters depend on others, such as refractive index and acoustic velocity on density. In our simulations we considered the former using the Lorentz-Lorenz equation:

$$n = \left(\frac{1+2R\rho}{1-R\rho}\right)^{1/2}, \quad (11)$$

where $R$ is the mass refraction (see Appendix B for definition). Thus, $n$ was varied both by the direct variation $\delta n$ and by the density variation $\delta \rho$ through Eq. (11). The dependences of other parameters on density including $v_{ac}$ are not so explicit in view of an inhomogeneity of the substance and a thermodynamic unsteadiness of the process, so that they were varied independently. The initial (unvaried) values of the parameters used in our simulation are collected in the Table II.

TABLE II. Physical properties of the solvents at $T = 293$ K and standard pressure

| Solvent | $n^{32,a}$ | $\rho$, kg m$^{-3}$ [32] | $v_{ac}$, m s$^{-1}$ [32] | $C_p$, J(kg K)$^{-1}$ | $\beta_T$, kK$^{-1}$ |
|---------|-----------|------------------------|---------------------------|----------------------|---------------------|
| NMP     | 1.471     | 1033                   | 1565                      | 1764[33,b]           | 0.83[34]            |
| Water   | 1.333     | 998                    | 1483                      | 4183[35]             | 0.20[35]            |

[a] At $\lambda = 590$ nm
[b] At $T = 298$ K

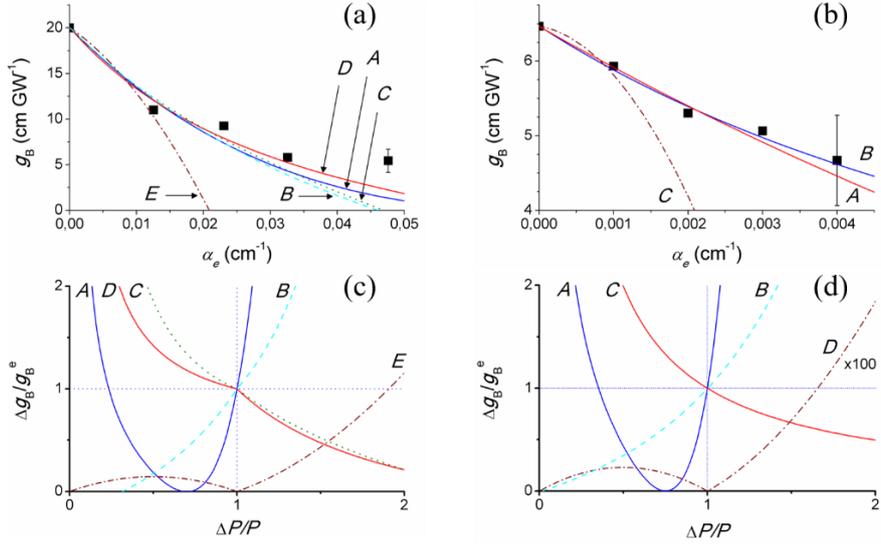

FIG. 8. SBS gain factor of NMP (a) and water (b) against linear absorption of graphene suspensions; points: experimental results; curves: simulations with variation of the parameters: (a) $A$, $\delta n$; $B$, $\delta\rho$; $C$, $\delta v_{ac}$; $D$, $\delta\Gamma_B$; $E$, $\delta\gamma_T$; (b) $A$, $\delta n$ and $\delta\rho$; $B$, $\delta v_{ac}$ and $\delta\Gamma_B$; $C$, $\delta\gamma_T$. Relative change of the SBS gain factor of NMP (c) and water (d) against variance of the parameters: (c) $A$, $\delta n$; $B$, $\delta\rho$; $C$, $\delta v_{ac}$; $D$, $\delta\Gamma_B$; $E$, $\delta\gamma_T$; (d) $A$, $\delta n$; $B$, $\delta\rho$; $C$, $\delta v_{ac}$ and $\delta\Gamma_B$; $D$, $\delta\gamma_T$.

Electrostrictive coefficients were also evaluated using Eq. (11), which gives a well-known relation: $\gamma_e = \varepsilon_0(n^2-1)(n^2+2)/3$. The Brillouin line widths $\Gamma_B$ ($\alpha_e = 0$) were taking those to fit the observed values of $g_B(0)$ in calculating by Eq. (5) at $\alpha_e = 0$: $\Gamma_B = 2.545$ GHz (NMP) and $\Gamma_B = 3.475$ GHz (water). The heat capacity variation $\delta C_p$ in $\gamma_T(\alpha_e)$ functionally gives the same behavior as $\delta\beta_T$ with the opposite sign, so that the latter can be considered as a variation of the thermo-optic coupling coefficient in total: $\delta\beta_T = \delta\gamma_T$.

The simulated $g_B(\alpha_e)$ dependences due to different parameter variations are superimposed onto experimental points in Figs. 8(a) and 8(b). The efficiency of each parameter towards the relative Brillouin gain change is illustrated by dependences in Figs. 8(c) and 8(d). The efficiency curves make it clear that the thermal expansion gives very small contribution to the Brillouin gain factor, which in case of water is only $ca.$ $10^{-3}$ of $g_B^e$ values in all the small variations range. In case of NMP, $g_B^T$ still has nonzero impact, which has been considered mostly in simulations at other parameter variations. The simulated curves of $g_B(\alpha_e)$ due to $\delta\gamma_T$ have too strong curvature [curve $E$ in Fig. 8(a) and curve $C$ in Fig. 8(b)] and cannot describe the effect observed.

In total, the simulation does not follow the experimental points sufficiently well. However, the proximity of each curve to them indicates the influence reliability of the corresponding parameter. Since the dependence is more nonlinear in case of NMP we look more closely at Figs. 8(a) and 8(c) as more indicative.

The variations of $v_{ac}$ and $\Gamma_B$, both positive (increase), give similar contribution to $g_B(\alpha_e)$, different in its value only for NMP because of the $g_B^T$ part which does not depend on $v_{ac}$. This circumstance allows to see on the example of NMP that the simulation due to $\delta v_{ac}$ [curve $C$ in Fig. 8(a)] gives a curvature different from that manifesting in the experiment. On the other hand, the positive variation $\delta\Gamma_B$ gives the simulated curve which approaches best to the experimental points. This makes $\Gamma_B$ more important parameter in the SBS quenching as compared to $v_{ac}$.

The negative variations of $\rho$ and $n$ (their decrease) give even stronger influence on the effect observed and reproduce the true sign of curvature [curves $A$, $B$ in Fig. 8(a) and curve $A$ in Fig. 8(b)]. The absorptive increment (decrements) providing the best fit of the experimental points are given in the Table III. The sensitivity of $g_B(\alpha_e)$ curves to these values defines their uncertainties which are within the last valid digits.

Lower $\Delta_\alpha P$ value show stronger influence of the corresponding parameter. Despite $n$ is the most influenced parameter, it is not independent, and its variation can be contributed from both density and mass refraction changes. It can be calculated that values $\Delta_\alpha(R\rho) = -11.1$ cm (NMP) and $-32$ cm (water) correspond to $\Delta_\alpha n$ indicated in the table. An analysis of the mass refraction impact given in Appendix II demonstrates the prevailing of density-determined nature of the refractive index changing.

Therefore, the simulation results denoted a density decrease and an increase of Brillouin line width as the key changes leading to the SBS quenching, whereas the effect is very low sensitive to the thermo-optic coupling. Analyzing parameters that determine $\Delta_\alpha\Gamma_B$ in Eq. (8), we can see that its growth is unlikely determined by the Brillouin frequency shift ($\omega_0 - \omega_B$), whose change should be rather negative upon irradiation.[29]

TABLE III. Relative absorptive changes (in cm) of refractive index, density, acoustic velocity, Brillouin line width and thermal optic coefficient providing the observed decrease of $g_B$

| Solvent | $\Delta_\alpha n$ | $\Delta_\alpha \rho$ | $\Delta_\alpha v_{ac}$ | $\Delta_\alpha \Gamma_B$ | $\Delta\gamma_T$ |
|---|---|---|---|---|---|
| NMP | $-4.4$ | $-14.8$ | 30 | 27 | 44 |
| Water | $-8.7$ | $-50$ | 99 | 99 | 2860 |

Thermodynamic studies show a remarkable and monotonous decrease of viscosity (both shear and bulk) of liquids[36] and steam[35] in a large range of temperature and pressure growth. It means that if the Eq. (8) remains adequate for the inhomogeneous media, the growth of $\Gamma_B$ should be attributed to a drop of acoustic velocity. Such kind of decrease is anticipated in a bubbly liquid due to a high compressibility of bubbles.[30] The effect can be considered as an acoustic wave extinction in growing bubbles that increase the acoustic absorption coefficient $\alpha_{ac} = \Gamma_B/v_{ac}$ of the liquid, which is most likely the main parameter in the SBS quenching along with density.

Apparently, both the density and acoustic absorption factors take place in the real process acting simultaneously. For this reason, then we performed a fitting of the experimental $g_B(\alpha_e)$ points by a simulation with simultaneous variation of $\rho$ and $\Gamma_B$ values. The curves obtained are shown in Figs. 4(c) and 4(d) and manifest a perfect coincidence with the measured dependence with correlation coefficients 0.992 (NMP) and 0.991 (water). Corresponding parameter changes are: $\Delta_\alpha\rho = -1.55$ cm (NMP), $-5.4$ cm (water) and $\Delta_\alpha\Gamma = 50$ cm (NMP), 90 cm (water).

### D. Bubble size scaling

Since the density reduction is determined by the bubble formation: $\delta\rho = \rho(1 - f_C)$, where $f_C$ is the bubble filling factor, the density decrement allows to estimate the average bubble radius:

$$r_b = \left(\frac{9d^2\sigma_e\rho_s\Delta_\alpha\rho}{4\pi}\right)^{1/3} \quad (12)$$

as it is shown in Appendix B. Here $d$ denotes the average flake characteristic size, and $\rho_s = 7.7\cdot10^{-8}$ g·cm$^{-2}$ [37] is the surface density of graphene. Estimating $d = 600$ nm from the AFM image scale, we can get the bubble radius: $r_b = 2.0 \pm 0.5$ μm. An error analysis is given in Appendix B. Since the SBS effect develops during few nanoseconds (due to the pulse shortening SBS intensity is negligible already after fourth nanosecond of the incident pulse) the radius obtained by this means corresponds to the starting point of bubble evolution: just after the termination of graphite sublimation. Thus, if to tune laser pulse duration and its energy, one can trace the bubble size evolution using the SBS method.

The bubble radius found is comparable with those obtained in SWCNT suspension in binary water-PEG solvent under nanosecond laser irradiation: 1.2 μm.[30] Therewith, the referenced study confirms a weak dependence of the initial bubble size on the laser pulse energy. Our one-point measurement of the SBS quenching in SWCNT suspension: $\Delta g_B/\Delta\alpha_e = -94$ cm$^2$GW$^{-1}$ is less than our result for graphene that will also give a lower value for the bubble size in case of nanotubes.

Simple evaluations following Eq. (B1) and Eq. (B2) and using the same density decrement value give the ratio of bubble-to-flake volume $V_b/V_f = 1\cdot10^5$ without data of flake surface size, but with the flake thickness. In our case it does not improve the uncertainty, but at some conditions the thickness can be made more uniform and precisely measured by AFM or Raman spectroscopy than the flake area. In this case the volume ratio can be more indicative.

In both cases the absorption cross-section of nanomaterial should be known. For graphene flakes numerous experiments give values which are several times different depending on the suspension preparation conditions.[38-41] And, despite similar $\sigma_e$ values were demonstrated for different organic solvents,[38] the value found in aqueous suspensions was twice less than that in NMP-based suspension after the same treatment procedure,[38,39] presumably, due to the surfactant influence. If, following this reason, we perform similar evaluations for NMP in assumption of the graphene absorption cross-section being twice lager than in water, we obtain values: $r_b = 1.7$ μm and $V_b/V_f = 8\cdot10^4$, which are comparable with the values obtained for aqueous suspension.

Despite the demonstrated possibility of the SBS method in obtaining the bubble size, the simultaneous fitting performed reveals a wide minimum which is crucial especially for $\Delta_\alpha\rho$ due to its small value and can increase the uncertainties of the characteristics found. Independent measurements of the acoustic absorption by photo-acoustic methods or Brillouin spectroscopy must be done to obtain the full image of the mechanism of light-matter interaction in conditions of the SBS type phase conjugation in liquid suspensions of strongly absorbing nanoparticles. Once the acoustic absorption coefficient is fixed, the dimension of vapor bubbles can be determined more precisely.

### V. CONCLUSION

In summary, we have studied the stimulated Brillouin scattering in NMP and water in the presence of nanoscale graphene flakes and have measured its energetic characteristics. We have found a strong effect of SBS quenching in liquids by graphene with such low concentrations of the nanoparticles that give no remarkable absorption in the material. Established linear dependences of SBS threshold on graphene absorption coefficient (concentration) can be used for the detection of small quantities of the nanomaterial in liquid media upwards of $5\cdot10^{-8}$ g cm$^{-3}$ which has been shown for the suspension in water.

Computer simulations of the Brillouin gain factor show the efficiency of different thermodynamic, electrooptic and photoacoustic parameters in the SBS quenching. The role of density and compressibility among them which change due to carbon vapor bubble formation is found to be decisive. Effective bubble size has been estimated, which can be specified in combination with acoustic wave absorption measurements providing more information on bubble properties in nanosecond time scale. From this point of view, the SBS method can be used as a tool for the nanosecond-resolved bubble formation scaling in nanoparticle contained media.

The studied effect can be used for the SBS suppression in situations where it is undesirable (both in laser technology and optical telecommunication networks). We expect an appearance of this effect in transparent solids containing traces of nanoparticles and believe that the method can be applied to other kinds of 2D-nanomaterials and maybe other nano-sized linear or nonlinear absorbers.

## ACKNOWLEDGEMENTS

The work is financially supported by the National Natural Science Foundation of China (61675217, 61522510); the Strategic Priority Research Program of Chinese Academy of Sciences (CAS) (XDB16030700); the Key Research Program of Frontier Science of CAS (QYZDB-SSW-JSC041); the Program of Shanghai Academic Research Leader (17XD1403900); President's International Fellowship Initiative (PIFI) of CAS (2017VTA0010, 2017VTB0006, 2018VTB0007).

## APPENDIX A: SUSPENSION PREPARATION AND ABSORPTION CROSS-SECTION MEASUREMENTS

Graphene suspensions were prepared from graphite powder (Sigma-Aldrich 332461) by the way like that described earlier.[16] NMP (99.5%, Sigma-Aldrich 328634) and double deionized water were used as solvents. Graphite powders were added to the solvent with initial concentration of 5 mg/ml. Sodium cholate (NaC) hydrate > 99% (Sigma-Aldrich C6445) surfactant was added to the aqueous sample with initial concentration 0.03 mg/ml. Mixtures were then sonicated using Sonics Vibra-cell VCX-750 W ultrasonic processor at the amplitude 38% for 1 h. Then the suspensions were centrifuged in Rotofix 32A centrifuge at 4000 rpm for 1 h twice with 15 h interval. After that supernatant portions were taken, and their absorption spectra were measured.

The precipitate was dried in an evacuated drying box at *ca.* 1 Torr pressure and T = 90°C with a periodical control of weight using Mettler Toledo XS105 analytic balance. The drying ended when the weight stopped changing. The total weight of the dry precipitate was subtracted from the initial weight to evaluate the carbon concentration $\rho_C$ in the suspension. For aqueous suspension we obtained $\rho_C$ = 56.3 ± 11.2 μg cm$^{-3}$. The uncertainty is mainly determined by the weight error ± 0.5 mg appearing from poorly controllable water vapor adsorption by dry carbon powder. No weight loss was found in carbon precipitate from NMP comparing to the initial carbon powder weight in several attempts of suspension preparation and drying during weeks. This apparently means that the procedure doesn't allow to eject NMP molecules adsorbed by carbon. After spectral measurements 0.47 mg cm$^{-3}$ of NaC was added to the aqueous suspension for better stability. The total concentration 0.5 wt.% of NaC in the suspension is close to its critical micelle concentration[19] so that we can expect that graphene flakes are totally surrounded by the surfactant. Both suspensions (NMP- and water-based) were stable, and no sedimentation was seen during more than 6 months.

Absorption spectra of the studied suspensions were measured in 1 cm path cell using PerkinElmer Lambda 750 dual-beam UV/VIS/NIR spectrometer, just after taking supernatant. The spectra were measured with reference to the solvent placed in the reference beam. The results are shown in Fig. 9.

The NMP suspension was approximately 5 times denser; its spectrum in the Fig. 9 is divided by 10 to compare with that of aqueous suspensions. Two main factors determine the absorption spectra shape. First, it is correlated with graphene density of states $DOS(\lambda)$, which in case of electron-hole symmetry is determined by the complete elliptic integral of first kind:[1]

$$DOS(\lambda) = \frac{8}{\pi^2 hc} \frac{\lambda_h^2}{\lambda} \sqrt{Z_0} \int_0^{\frac{\pi}{2}} \frac{d\phi}{Z_0 - Z_1 \sin^2(\phi)},$$

$$Z_0 = \left(1 + \frac{\lambda_h}{\lambda}\right)^2 + \frac{1}{4}\left[\left(\frac{\lambda_h}{\lambda}\right)^2 - 1\right]^2, \quad \text{(A1)}$$

$$Z_1 = 4\frac{\lambda_h}{\lambda}.$$

Here $\lambda_h$ is the peak position of $DOS$ which corresponds to twofold nearest-neighbor hopping energy in graphene ($E_h \approx 2.8$ eV), $h$ is the Plank constant, $c$ – the speed of light in vacuum. The calculated $DOS(\lambda)$ function is shown in the Fig. 9 by a dotted curve with maximum at $\lambda_h$ = 221.4 nm. The other dotted curve with maximum at $\lambda_h$ = 270 nm is a Lorentz function which is a good approximation for a surface plasmon absorption cross-section.[42] The transverse plasmon resonance frequency in graphene takes place at ~5 eV (see Supplementary material to Ref. 43).

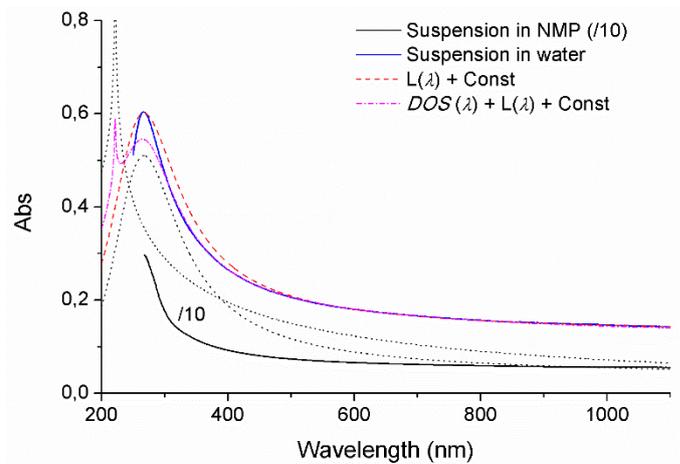

FIG. 9. Absorption spectra of graphene suspensions in water and NMP as prepared.

It can be seen in the Fig. 9 on the example of aqueous suspension spectrum, that each of these dependences do not describe it separately. The Lorentzian can describe only its red part for wavelengths $\lambda > 500$ nm with a constant adduct (red dashed curve). Both *DOS* and Lorentzian together reproduce the experimental curve down to $\lambda \approx 300$ nm (magenta dash-dotted curve). The constant adduct is also necessary to fit the spectrum, and it is apparently connected with the linear Dirac electronic spectrum of graphene, which leads to the constant absorption coefficient equal only $\alpha\pi$ per layer, where $\alpha = 1/137$ is the fine structure constant, in all NIR and visible spectral region down to 450 nm.[43]

The peak shape in Fig. 9 may be strongly influenced by light scattering which is small in visible for the nanoparticles concentrations studied, but becomes remarkable in UV not only due to diffraction enhancement, but also due to dramatic changes in dynamic dielectric permittivity through the factor $(\varepsilon(\omega) - \varepsilon_{NaC}(\omega))$.

These considerations assume a complicated character of the spectrum determined by different electron movements in graphene. In a real graphene flake electron-hole symmetry can be broken due to defects and a confinement effect that leads to a shift of the *DOS* peak position and distortion of its shape. The real spectrum of graphene suspension is even more complicated being a sum of different particles absorption, where the size and layer thickness distribution as well as defect structures are crucial.

Thus, the total absorption coefficient of graphene in water measured at $\lambda = 532$ nm is $\alpha_{10} = 0.1954 \pm 0.0011$ cm$^{-1}$ that taking into account $\rho_C$ value obtained gives the absorption cross-section $\sigma_e = \alpha_{10} \cdot \ln(10)/\rho_C = (0.80 \pm 0.16) \cdot 10^4$ cm$^2$g$^{-1}$. It is a little less than the value $1.39 \cdot 10^4$ cm$^2$g$^{-1}$ obtained for low-concentrated aqueous graphene suspensions,[39] which we explain by difference in layer thickness and defect structures of our graphene flakes compared to those studied in the reference.

## APPENDIX B: CARBON FILLING FACTORS AND MASS REFRACTION

Fast carbon evaporation during 4 ns pulse leads to an increase of total volume of bubbles $V_b$. The bubble filling factor, which is the ratio of its volume to the caustic volume can be then evaluated: $f_C = V_b/V = r_b^3 N_b/(2Z_R W_0^2)$, where $r_b$ is the bubble radius and $N_b$ is the number of bubbles in the caustic region. The latter coincides with the number of graphite flakes in the same volume. We can calculate it from carbon density $\rho_C$, $V$, and the mass of one particle $m_{Gr}$, which can be obtained only by rough estimations. Assuming three-layer thickness in average, we can accept: $m_{Gr} = 3d^2\rho_s$, where $\rho_s = 7.7 \cdot 10^{-8}$ g·cm$^{-2}$ is the surface density of graphene.[37] Then, considering that $\alpha_e = \rho_C \sigma_e$, we obtain:

$$N_b = \frac{8\pi W_0^2 Z_R \alpha_e}{9d^2 \sigma_e \rho_s},$$

which gives the value $N_b = 1.5 \cdot 10^6 \cdot \alpha_e$(cm$^{-1}$) with the assumption of the average characteristic flake size based on the AFM image scale to be around $d = 600$ nm. Its uncertainty comes mainly from the size distribution standard deviation SD($d$), and apparently consists a few tens of percent. Thus, for the bubble filling factor we have:

$$f_C = \frac{4\pi \cdot r_b^3 \alpha_e}{9d^2 \sigma_e \rho_s},$$

or $f_C = 5 \cdot 10^{11} \cdot [r_b(\text{cm})]^3 \cdot \alpha_e(\text{cm}^{-1})$ in water with the said uncertainty.

At that, the initial graphene flakes filling factor $f_0$ can be estimated more precisely because it does not require the particle size, but only its thickness h:

$$f_0 = \frac{h \cdot \alpha_e}{3\sigma_e \rho_s}. \tag{B1}$$

Assuming the middle thickness of graphene flakes in our suspension h = 1 nm, we obtain $f_0 = 5.4 \cdot 10^{-5} \cdot \alpha_e$(cm$^{-1}$) for water. At the average number of layers equal to three, its standard deviation cannot be more than one that implies the 33% upper estimation for the $\Delta$h uncertainty. With the 20% uncertainty of density $\rho_C$ it gives $\Delta f_0/f_0 = 39\%$.

From the definition of filling factors also follows that their ratio is just a ratio of the average bubble volume to the average flake volume:

$$\frac{f_C}{f_0} = \frac{V_b}{V_f}. \tag{B2}$$

Mass refraction is a ratio of the Lorentz-Lorenz function to the mass density:

$$R \equiv \frac{n^2-1}{n^2+2}\frac{1}{\rho}, \tag{B3}$$

and it is a measure of polarizability $a_0$ of a particle with the mass $M_0$:

$$R = \frac{4\pi}{3}\frac{a_0}{M_0} \tag{B4}$$

In application to the substances under study, $R$ was determined from Eq. (B3) for NMP: $R = 0.271$ cm$^3$g$^{-1}$ and for water: $R = 0.206$ cm$^3$g$^{-1}$ using data presented in Table II. For the graphene mass refraction, a refractive index value $n_{3layer} = 2.27$ for the three-layered graphene,[44] and a graphene density simple estimation: $\rho_{Gr} = 3\rho_s/h = 2.31$ g cm$^{-3}$ were used: $R_{Gr} = 0.251$ cm$^3$g$^{-1}$.

Refractive index of the suspension is not practically affected by the mass refraction of graphene in view of its very low density:

$$n_{susp} = \left[\frac{1+2(R_{solv}\rho_{solv}+R_{Gr}\rho_C)}{1-(R_{solv}\rho_{solv}+R_{Gr}\rho_C)}\right]^{1/2},$$

$n_{susp}$ = 1.472 (suspension in NMP) and 1.333 (suspension in water) at maximal concentrations used.

The mass refraction of carbon in bubbles is evaluated from Eq. (B4) and literature data on carbon atom polarizability:[45] $R_C$ = 0.374 cm$^3$g$^{-1}$ and found even greater than all other refractions. Thus, carbon vaporization could increase the refractive index in changing the term $R_{Gr}\rho_C$, which is negligible, by $R_C\rho_C/f_C = R_C/(5\cdot10^{11}\cdot r_b^3\sigma_e)$ which can be larger than solvent's ones at small bubbles. Since we know from the experiment and related computer simulations that the refractive index of the suspension heated by laser pulse is decreasing, we can establish the conditions on the bubble radius value: $r_b \gg$ 10 nm. In that case $R_C\rho_C/f_C \ll 0.1$, and the decreasing of the refraction is assured only by the solvent density reduction due to bubbles formation:

$$\delta(R\rho) = R_{solv}\rho_{solv}(1-f_C). \quad (B5)$$

From here we see that $f_C = \Delta_\alpha\rho\cdot\Delta\alpha_e$, where $\Delta_\alpha\rho$ the absorptive decrement of density introduced by Eq. (10). The substitution of Eq. (B1) into Eq. (B2) and then into Eq. (B5) gives us the following expression for the bubble size:

$$r_b = \left(\frac{9d^2\sigma_e\rho_s\Delta_\alpha\rho}{4\pi}\right)^{1/3}. \quad (B6)$$

The expression is exact, *i.e.*, obtained from the geometric consideration with the only assumption of the square shape of the flake. In case of other shape types $d^2$ should be replaced by the corresponding area value. Therefore, the bubble radius depends on the flake surface area, absorption cross-section and absorptive decrement of density.

Roughly supposing the relative standard deviation of the flake size distribution to be $SD(d)/d = 30\%$, the fractional error $\delta_d r_b$ due to flake size uncertainty is 23%. The 20% uncertainty of the absorption cross-section gives the fractional error $\delta_\sigma r_b$ = 7%. The error in $\Delta_\alpha\rho$ obtained from the experimental data fitting is not large and consists only few percent, however, other minima in the fitting related to other $\Delta_\alpha\Gamma$ values are possible. In total, we evaluate $\delta r_b$ error by 25%. Therefore, simultaneous measurement of acoustic absorption will help to obtain more reliable data of average bubble size.